\documentclass[twocolumn,aps,pra,amsmath,amssymb,superscriptaddress]{revtex4-1}

\usepackage[usenames]{color}
\usepackage{hyperref}
\usepackage{graphicx}

\def\<{\langle}
\def\>{\rangle}

\begin{document}

\title{Two-body and three-body contacts for three bosons in the unitary regime:  Analytic expressions and limiting forms}

\author{V. E. Colussi}
\email[Corresponding author:  ]{colussiv@gmail.com}
\affiliation{Eindhoven University of Technology, PO Box 513, 5600 MB Eindhoven, The Netherlands}

\begin{abstract}
The two and three-body contacts are central to a set of univeral relations between microscopic few-body physics within an ultracold Bose gas and its thermodynamical properties.  They may also be defined in trapped few-particle systems, which is the subject of this work.  In this work, we focus on the unitary three-body problem in a trap, where interactions are as strong as allowed by quantum mechanics.  We derive analytic results for the two and three-body contacts in this regime and compare with existing limiting expressions and previous numerical studies.
\end{abstract}

\maketitle

\section{Introduction}
As the s-wave scattering length approaches unitarity $|a|\to\infty$, properties of both macroscopic and microscopic strongly-interacting ultracold quantum systems can be described effectively by a reduced set of remaining finite quantities.   For three trapped bosonic alkali atoms at unitarity, physics in the unitary regime is parametrized solely by the range of the potential, captured by the van der Waals length $r_\mathrm{vdW}$ \cite{RevModPhys.82.1225}, and the trap length $a_\mathrm{ho}=(\hbar/m\omega)^{1/2}$, with frequency $\omega$ and single-particle mass $m$.  Consequently, these scales parametrize the two and three-body contacts, respectively, that are central to many relevant observables in the system including the tail of the single-particle density, short-distance behavior of correlation functions, high-frequency tail of the rf transition rate, virial theorem, and total energy of the system \cite{PhysRevA.86.053633,PhysRevLett.106.153005}.  Intuitively, the two and three-body contacts measure the mean number of clustered pairs and triplets of bosons, respectively \cite{PhysRevA.78.053606}.  The centrality of the extensive two-body contact $C_2$ to the thermodynamic properties of the two-component unitary Fermi gas was worked out by Shina Tan~\cite{TAN20082952,TAN20082971,TAN20082987}, and this quantity frames part of our current understanding of experiments in this regime~\cite{zwerger2011bcs}.  For strongly-interacting Bose gases, however, the presence of the Efimov effect \cite{efimov1971weakly,d2017few} and associated infinite sequence of bound trimers requires the introduction of an additional parameter, the extensive three-body contact $C_3$ \cite{PhysRevA.86.053633,PhysRevLett.106.153005}.  In the strongly-interacting regime, where interaction length scales are diverging, the contacts may also be used to interpolate in between perturbative results in neighboring regimes \cite{PhysRevLett.118.103403}.

The centrality of the contacts means that they may be measured through various means.  By measuring the high-momentum tail of the rf transition rate in a weakly-interacting Bose gas, Wild {\it et al.} \cite{PhysRevLett.108.145305} measured a nonzero $C_2$ and found that $C_3$ was consistent with zero.  In the non-equilibrium regime of a nondegenerate Bose gas quenched to unitarity, both $C_2$ and $C_3$ were measured interferrometrically in Ref.~\cite{Fletcher377}, finding a gradual approach of $C_3$ to equilibrium.  It has also been suggested that the high-momentum tail  of the degenerate Bose gas quenched to unitarity measured in Ref.~\cite{makotyn2014universal} is a nonzero measurement of $C_3$ \cite{PhysRevLett.112.110402}, although there is currently no consensus and a recent experimental study by Eigen {\it et al.} \cite{eigen2018prethermal} found no evidence of the contacts in the single-particle momentum distribution at early-times after the quench.  Theoretically, a prediction for the gradual growth of $C_3$ in a uniform degenerate Bose gas quenched to unitary was made in Ref.~\cite{PhysRevLett.120.100401} where a coherent beating phenomenon at the frequency of Efimov states was observed. 

 The contacts can also be measured in microscopic few-body systems for instance by measuring interaction energies or spectroscopically.  Theoretically, the contacts corresponding to the lowest few eigenstates were calculated numerically for three and four trapped bosons in Ref.~\cite{PhysRevA.97.033621} at fixed trap length.  In the present work, we provide analytic results for $C_2$ and $C_3$ for the trapped unitary three-body problem including analytic expressions and limiting forms.  This includes a review of some of previous results derived already in the supplementary materials of Ref.~\cite{PhysRevLett.120.100401} for $C_3$, included here in more detail, in addition to new derivations for $C_2$.  

In this work, we begin in Sec.~\ref{sec:background} by reviewing analytic solutions of the unitary three-boson problem in a trap in the zero-range model first derived in Ref.~\cite{PhysRevLett.97.150401}.  In Sec.~\ref{sec:c3}, we use a relation between the short-distance behavior of the three-body correlation function and $C_3$ from Ref.~\cite{PhysRevLett.120.100401} to derive analytic expressions for the three-body contacts for each trapped three-body eigenstate including limiting expressions in the large-energy limit.  In Sec.~\ref{sec:c2}, we perform an analogous derivation except for the two-body contacts for each trapped three-body eigenstate.  Finally, in Sec.~\ref{sec:discussion}, we discuss our findings and compare against the available existing results in \cite{PhysRevA.97.033621} which evaluated $C_2$ and $C_3$ including more realistic finite-range effects at fixed trap length.  We find generally excellent agreement with the available results of that work.


\section{Three trapped bosons at unitarity}\label{sec:background}
In this section, we outline analytic solutions of the unitary three-body problem in a trap given in Ref.~\cite{PhysRevLett.97.150401}.   Consider three non-interacting bosons located at ${\bf r}_1,\ {\bf r}_2,\ {\bf r}_3$ in a harmonic trap whose wave function $\Psi({\bf r}_1, {\bf r}_2, {\bf r}_3)$ solves the non-interacting Schrödinger equation
\begin{equation}\label{eq:3bschrodinger}
\sum_{l=1}^3\left[-\frac{\hbar^2}{2m}\nabla^2_{{\bf r}_l}+\frac{1}{2}m\omega^2 r_i^2\right]\Psi=E\Psi.
\end{equation}
The three-body wave function can be separated $|\Psi\rangle=|\Psi_\mathrm{CM}\rangle|\psi\rangle$ into center of mass $|\Psi_\mathrm{CM}\rangle$ and relative $|\psi\rangle$ components.  The center of mass coordinate is defined as ${\bf C}\equiv({\bf r}_1+{\bf r}_2+{\bf r}_3)/3$, and the relative motion is parametrized by the hyperradius $R^2\equiv(r^2+\rho^2)/2$ and hyperangles ${\bf \Omega}\equiv\{\hat{\rho},\hat{r},\alpha\}$ where ${\bf r}\equiv{\bf r}_2-{\bf r}_1$ and $\boldsymbol{\rho}=(2{\bf r}_3-{\bf r}_1-{\bf r}_2)/\sqrt{3}$ are the Jacobi vectors with spherical angles $\hat{\rho}$ and $\hat{r}$, respectively.  The hyperangle $\alpha$ is defined as $\tan\alpha=r/\rho$, and is related to each individual Jacobi vector as $r=\sqrt{2}R\sin\alpha$, $\rho=\sqrt{2}R\cos\alpha$ where $\alpha\in[0,\pi/2]$.  

In ultracold quantum gases of alkali atoms, it is generally the case that the typical momentum of the problem are such that $kr_\mathrm{vdW}\ll1$, where $r_\mathrm{vdW}$ is the van der Waals length which furnishes a natural short-range for pairwise interactions \cite{RevModPhys.82.1225}.  In this limit, the zero-range approximation can be made, and pairwise interactions can be accounted for via the Bethe-Peierls boundary conditions
\begin{equation}\label{eq:bethe}
\psi(R,{\bf \Omega}))\underset{{|\bf r|}\rightarrow 0}{=}\left(\frac{1}{a}-\frac{1}{r}\right)A(\boldsymbol{\rho}),
\end{equation}
where the unknown function $A$ is determined by solving the three-body Schrödinger equation (Eq.~\eqref{eq:3bschrodinger}) with the above boundary condition.

In the unitary regime, $|a|\to\infty$ in Eq.~\eqref{eq:bethe} and the relative three-body wave function becomes singular as $1/r$ in the $r\to0$ limit.  Following Ref.~\cite{PhysRevLett.97.150401} and the original approach of Efimov \cite{efimov1971weakly}, the relative three-body wave function in this regime can be decomposed as $\psi(R,{\bf \Omega})=R^{-2}\mathcal{N}F(R)\phi({\bf \Omega})$ in terms of hyperradial $F(R)$ and hyperrangular $\phi({\bf \Omega}))=(1+\hat{Q})\varphi(\alpha)/\sin2\alpha\sqrt{4\pi}$ functions with normalization constant $\mathcal{N}$.  The operator $\hat{Q}=\hat{P}_{13}+\hat{P}_{23}$ is written in terms of the permuation operator $\hat{P}_{ij}$ that swaps particles $i$ and $j$.  The hyperangular basis functions $\varphi$ are determined from solutions of the equations \cite{PhysRevLett.97.150401}
\begin{eqnarray}
&&-\varphi_s''(\alpha)=s^2\varphi_s(\alpha),\label{eq:ode1}\\
&&\varphi_s(\pi/2)=0,\label{eq:ode2}\\
&&\varphi_s'(0)+\frac{8}{\sqrt{3}}\varphi_s(\pi/3)=0,\label{eq:ode3}
\end{eqnarray}
which are solved by $\varphi_s(\alpha)=\sin(s(\pi/2-\alpha))$.  The channel index $s$ is obtained by solving the transcendental equation
\begin{equation}\label{eq:trans}
\frac{8}{\sqrt{3}}\sin\left(\frac{s\pi}{6}\right)-s\cos\left(\frac{s\pi}{2}\right)=0.
\end{equation}
Each channel supports an associated set of hyperradial eigenfunctions $F^{(s)}_j(R)$ satisfying
\begin{widetext}
\begin{equation}
\left[-\frac{\hbar^2}{2m}\left(\frac{d^2}{dR^2}+\frac{1}{R}\frac{d}{dR}\right)+\frac{\hbar^2s^2}{2mR^2}+\frac{m\omega^2R^2}{2}\right]F_j^{(s)}(R)=E_\mathrm{3b}^{(s,j)}F^{(s)}_j(R),\label{eq:hyper}
\end{equation}
\end{widetext}
with quantum number $j=0,1,2,\dots$ and relative three-body eigenenergy $E_\mathrm{3b}^{(s,j)}$.  

The transcendental equation [Eq.~\eqref{eq:trans}] has both real and imaginary solutions.  For $s^2>0$, Eq.~\eqref{eq:hyper} has the solution
\begin{equation}\label{eq:funiversal}
F_j^{(s)}(R)=e^{-R^2/2a_\mathrm{ho}^2}L_j^{(s)}(R^2/a_\mathrm{ho}^2)\left(\frac{R}{a_\mathrm{ho}}\right)^s,
\end{equation}
where $L_j^{(s)}$ is a generalized Laguerre polynomial of degree $j$, and the associated eigenenergies are $E_\mathrm{3b}^{(s,j)}=\hbar\omega(s+2j+1)$.  We label the channels with $s^2>0$ as universal because they depend solely on the trapping parameters and not on microscopic details of the interaction.

The lone imaginary root of Eq.~\eqref{eq:trans} $s=is_0$ denotes the Efimov channel, where $s_0\approx 1.00624$ is Efimov's constant  \cite{efimov1971weakly}.  Equation~\eqref{eq:hyper} must be supplemented with the additional boundary condition $F_j^{(s_0)}(R)\propto \sin[s_0\ln(R/R_t)]$ to preserve the Hermiticity of the problem \cite{danilov1961gs,PhysRevLett.97.150401}, where $R_t$ is a three-body parameter setting the phase of the log-periodic oscillation.  The hyperradial solutions in this channel are
\begin{equation}\label{eq:whittaker}
F_j^{(s_0)}(R)=\frac{a_\mathrm{ho}}{R}W_{E^{(s_0,j)}_\mathrm{3b}/2\hbar\omega,is_0/2}(R^2/a_\mathrm{ho}^2),
\end{equation}
where $W$ is a Whittaker function \cite{abramowitz1964handbook}.  The eigenenergy spectrum in the Efimov channel is obtained by solving
\begin{align}
\arg\Gamma&\left[1+is_0\right]\nonumber\\
&=\arg\Gamma\left[\frac{1+is_0-E_\mathrm{3b}^{(s_0,j)}/\hbar\omega}{2}\right]+s_0\ln\frac{R_t}{a_\mathrm{ho}}\label{eq:spectrum}
\end{align}
which is understood $\mod \pi$ and was first obtained in Ref.~\cite{PhysRevLett.89.250401}.  To make the connection between the three-body parameters $R_t$ and $\kappa_*$, we take the $j\to-\infty$ limit of Eq.~\ref{eq:spectrum}, using the relations \cite{abramowitz1964handbook}
\begin{align}
\arg\Gamma&\left[\frac{1+is_0-E_\mathrm{3b}^{(s_0,j)}/\hbar\omega}{2}\right]\nonumber\\
&=\frac{s_0}{2}\psi\left(\frac{1-E_\mathrm{3b}^{(s_0,j)}/\hbar\omega}{2}\right),\nonumber\\
&\approx \frac{s_0}{2}\ln\left(\frac{-E_\mathrm{3b}^{(s_0,j)}/\hbar\omega}{2}\right),
\end{align}
where $\psi$ in the above equation is the digamma function.
Inserting this limiting result into Eq.~\eqref{eq:spectrum} gives the free-space result for the binding energies of Efimov trimers
\begin{equation}\label{eq:jnegative}
E_\mathrm{3b}^{(s_0,j)}\underset{{j}\rightarrow -\infty}{=}\frac{-2\hbar^2}{mR_t^2}e^{\frac{2}{s_0}(\arg\Gamma(1+is_0)-\pi j)}.
\end{equation}
Setting $j=0$ in the above equation gives $R_t=\sqrt{2}\exp(\text{Im}\ln[\Gamma(1+is_0)]/s_0)/\kappa_*$ in terms of the wave number of the $j=0$ trimer $E_\mathrm{3b}^{(s_0,0)}=-\hbar^2\kappa_*^2/m$.  We now choose the three-body parameter $R_t$ such that $E_\mathrm{3b}^{(s_0,0)}$ matches the ground-state trimer energy at unitarity for neutral atoms studied in Ref.~\cite{PhysRevLett.108.263001}.  In that work, the universality of $\kappa_*$ with the van der Waals length was established, finding $\kappa_*=0.226/r_\mathrm{vdW}$.  Therefore, we make the arbitrary distinction that $j=0$ is the ``ground-state'' Efimov trimer in the sense of Ref.~\cite{PhysRevLett.108.263001} and $j<0$ comprises the remainder of the Thomas collapse \cite{PhysRev.47.903}.  Taking the $j\to-\infty$ limit of the Whittaker function in Eq.~\ref{eq:whittaker} gives
\begin{equation}\label{eq:psiefimov}
F_j^{s_0}(R)\underset{{j}\rightarrow -\infty}{=}K_{is_0}(\sqrt{2}R\sqrt{|E_\mathrm{3b}^{(s_0,j)}|/\hbar\omega}/a_\mathrm{ho}),
\end{equation}
where $K_{is_0}$ is a modified Bessel function of imaginary argument, which is the free-space bound-state wave function for an Efimov trimer \cite{efimov1971weakly}.  The Efimov channel depends on the trapping parameters and also on $R_t$, which is sensitive to the microscopic length scale $r_\mathrm{vdW}$.  The Efimov channel is therefore not universal.  Furthermore, although $R_t$ depends on $r_\mathrm{vdW}$, the model remains zero-range involving only contact potentials.  We contrast this with the approach of Ref.~\cite{PhysRevA.97.033621}, which used more realistic finite-ranged potentials.  

For completeness, we discuss also the $j\to+\infty$ limit of Eq.~\ref{eq:spectrum}, where Stirling's formula gives
\begin{align}
&\frac{s_0}{2}\ln(E_\mathrm{3b}^{(s_0,j)}/\hbar\omega)-\frac{\pi E_\mathrm{3b}^{(s_0,j)}}{2\hbar\omega}\nonumber\\
&\quad\quad\quad\approx\arg\Gamma\left[\frac{1+is_0-E_\mathrm{3b}^{(s_0,j)}/\hbar\omega}{2}\right].
\end{align}
Therefore, Eq.~\ref{eq:spectrum} in this limit becomes
\begin{align}
E_\mathrm{3b}^{(s_0,j)}/\hbar\omega=&2j+\frac{1}{\pi}\left(s_0\ln(R_t^2/a_\mathrm{ho}^2)-2\arg\Gamma(1+is_0)\right)\nonumber\\
&+\frac{s_0}{\pi}\ln(E_\mathrm{3b}^{(s_0,j)}/\hbar\omega),\label{eq:jbig}
\end{align}
which is in general difficult to solve.  However, Eq.~\ref{eq:jbig}  is written in a form ``$f(x)=a+g(f(x))$'' that can be iterated.  In the $E_\mathrm{3b}^{(s_0,j)}\to\infty$ limit this produces corrections in powers of $j$ as
\begin{align}
E_\mathrm{3b}^{(s_0,j)}/\hbar\omega\underset{{j}\rightarrow +\infty}{=}&\frac{1}{\pi}\left(s_0\ln(jR_t^2/a_\mathrm{ho}^2)-2\arg\Gamma(1+is_0)\right)\nonumber\\
&+2j+O(j^{-1}\ln j),\label{eq:jpos}
\end{align}
which was first obtained in Ref.~\cite{wernerthesis}.  

The normalization constants $\mathcal{N}_{s,j}$ are defined as
\begin{equation}
\mathcal{N}_{s,j}^2\equiv\left(\frac{2}{\sqrt{3}}\right)^3\frac{1}{\langle F_j^{(s)}|F_j^{(s)}\rangle\langle\phi_s|\phi_s\rangle},\label{eq:norm}
\end{equation}
where the hyperangular inner product is defined in terms of the hyperangular solid angle as 
\begin{widetext}
\begin{align}
\langle \phi_s|\phi_s\rangle&\equiv\int d{\bf \Omega} |\phi_s({\bf \Omega})|^2,\\
&=-\frac{12\pi}{s}\sin\left(\frac{s^*\pi}{2}\right)
\left[\cos\left(\frac{s\pi}{2}\right)-\frac{s\pi}{2}\sin\left(\frac{s\pi}{2}\right)-\frac{4\pi}{3\sqrt{3}}\cos\left(\frac{s\pi}{6}\right)\right],\label{eq:hyperangleinner}\\
\int d{\bf \Omega}&\equiv2\int d\alpha\sin^2 2\alpha\int_0^{\pi/2} d\hat{\boldsymbol{\rho}}\int d\hat{\bf r}.
\end{align}
\end{widetext}
The result in Eq.~\ref{eq:hyperangleinner}, was first calculated in Ref.~\cite{wernerthesis} using a change of variables trick due to Efimov in Ref.~\cite{PhysRevC.47.1876}.  The hyperradial inner product is defined as
\begin{widetext}
\begin{align}
\langle F_j^{(s)}|F_{j}^{(s)}\rangle&\equiv\int_0^\infty dR\ R |F_j^{(s)}(R)|^2,\\
 \langle F_j^{(s)}|F_j^{(s)}\rangle&=a_\mathrm{ho}^2\frac{\Gamma\left[s+1+j\right]}{2j!}\quad\quad(s^2>0),\\
  \langle F_j^{(s_0)}|F_j^{(s_0)}\rangle&=a_\mathrm{ho}^2\frac{\pi\cdot \mathrm{Im} \psi\left(\frac{1-E_\mathrm{3b}^{(s_0,j)}/\hbar\omega+is_0}{2}\right)}{\sinh(s_0\pi)\cdot\left|\Gamma\left(\frac{1-E_\mathrm{3b}^{(s_0,j)}/\hbar\omega+is_0}{2}\right)\right|^2},\\
   \langle F_j^{(s_0)}|F_j^{(s_0)}\rangle&\underset{{j}\rightarrow-\infty}{=}a_\mathrm{ho}^2\frac{\pi s_0}{4|E_\mathrm{3b}^{(s_0,j)}/\hbar\omega|\sinh(\pi s_0)},   \label{eq:ffefimov}
\end{align}
\end{widetext}
where the relevant integrals can be found tabulated in Ref.~\cite{gradshteyn2007table}.  The factor $(2/\sqrt{3})^3$ in Eq.~\eqref{eq:norm} is due to the Jacobian ($D$) of the transformation $D({\bf r}_1,{\bf r}_2,{\bf r}_3)/D({\bf C},R,{\bf \Omega})=(\sqrt{3}/2)^3$, required so that $\langle\Psi|\Psi\rangle=\langle \Psi_\mathrm{CM}|\Psi_\mathrm{CM}\rangle\langle \psi_{s,j}|\psi_{s,j}\rangle=1$.

\section{Three-body contact $C_3$}\label{sec:c3}
Any one of the universal relations involving the contacts can be used as a starting point, however, we choose to obtain them through the limiting behavior of few-body correlation functions \cite{PhysRevLett.106.153005,PhysRevA.86.053633}.  We begin by deriving the extensive three-body contacts $C_3$ for each trapped three-body eigenstate.  

The non-normalized triplet correlation function is defined in second-quantization as \cite{cohen2011advances}
\begin{equation}\label{eq:g3def}
G^{(3)}({\bf r}_1,{\bf r}_2,{\bf r}_3)\equiv\langle \hat{\psi}^\dagger({\bf r}_1)  \hat{\psi}^\dagger({\bf r}_2)  \hat{\psi}^\dagger({\bf r}_3)   \hat{\psi}({\bf r}_3) \hat{\psi}({\bf r}_2)  \hat{\psi}({\bf r}_1)\rangle,
\end{equation}
in terms of the bosonic field operators $\hat{\psi}({\bf r})=\sum_{\bf k}\hat{a}_{\bf k}e^{i{\bf k}\cdot{\bf r}}/\sqrt{V}$ and system volume $V$.  For three bosons in vacuum, this is equivalent in first quantization to
\begin{equation}
G^{(3)}({\bf r}_1,{\bf r}_2,{\bf r}_3)=3!|\Psi({\bf r}_1,{\bf r}_2,{\bf r}_3)|^2.
\end{equation}
Following \cite{PhysRevA.86.053633}, we integrate out the center of mass dependence, and take the limit $R\to0$ at fixed ${\bf \Omega}$
\begin{align}  
 3!\int d^3 C\ \Psi({\bf r}_1,{\bf r}_2,{\bf r}_3)&=3!|\psi_{s,j}(R,{\bf \Omega})|^2,\nonumber\\
 &\underset{{R}\rightarrow 0}{=}|\psi_{sc}(R,{\bf \Omega})|^2\frac{8}{\sqrt{3}s_0^2}C_3^{(s,j)},\label{eq:g3c3}
 \end{align}
 for the j$^\text{th}$ eigenstate in the Efimov channel.  In the above equation, we have indexed the extensive three-body contact specific to the eigenstate $(s,j)$ as $C^{(s,j)}$.  In Eq.~\eqref{eq:g3c3}, $\psi_{sc}(R,{\bf \Omega})$ is zero-energy three-body scattering state \cite{PhysRevA.86.053633}
\begin{equation}
\psi_{sc}(R,{\bf \Omega})=\frac{1}{R^2}\sin\left[s_0\ln\frac{R}{R_t}\right]\frac{\phi_{s_0}({\bf \Omega})}{\sqrt{\langle\phi_{s_0}|\phi_{s_0}\rangle}}.
\end{equation}
Integrating both sides of Eq.~\eqref{eq:g3c3} over $\int d{\bf \Omega}$, and taking advantage of the orthogonality of the hyperangular eigenfunctions so that only $s=is_0$ need be considered yields
 \begin{equation}
\frac{s_0^2}{4\langle F_j^{(s_0)}|F_j^{(s_0)}\rangle}\left|\frac{F_j^{(s_0)}(R)}{\sin[s_0\ln R/R_t]}\right|^2\underset{{R}\rightarrow 0}{=}C_3^{(s_0,j)}.
\end{equation}
From the asymptotic behavior of the Whittaker functions \cite{abramowitz1964handbook}, the $R\to0$ limit in the above equation can be taken with the simple result
 \begin{align}
 a_\mathrm{ho}^2C_3^{(s_0j)}&=\frac{s_0}{ \mathrm{Im} \psi\left(\frac{1-E_\mathrm{3b}^{(s_0,j)}/\hbar\omega+is_0}{2}\right)}.\label{eq:c3trap}
 \end{align}
In the above equation, $C_3$ is made dimensionless by rescaling in powers of the oscillator length.  In the following two subsections, we derive first the $|E_\mathrm{3b}|/\hbar\omega\to\infty$ limiting forms of Eq.~\eqref{eq:c3trap} and then show some results at intermediate energies.

\subsection{Limiting cases}\label{sec:c3limiting}
In this subsection, we derive the $|E_\mathrm{3b}|/\hbar\omega\to\infty$ limiting forms of Eq.~\eqref{eq:c3trap}.  

{\bf Case I  $(E_\mathrm{3b}/\hbar\omega\to-\infty)$:}  In this limit, the digamma function has the asymptotic form \cite{abramowitz1964handbook}
\begin{equation}\label{eq:psilarge}
\psi(z)=\ln z+O(z^{-1})\quad\quad (z\to\infty\text{ with }|\arg z|<\pi).
\end{equation}
Using the property $\ln z=\ln|z|+i\arg z$, we find
\begin{align}
\mathrm{Im}\ \psi&\left(\frac{1-E^{(s_0,j)}_\mathrm{3b}/\hbar\omega+is_0}{2}\right)\nonumber\\
&\quad=\arg\left(1-E^{(s_0,j)}_\mathrm{3b}/\hbar\omega+is_0\right),\nonumber\\
&\quad=\arctan\left(\frac{s_0}{1-E^{(s_0,j)}_\mathrm{3b}/\hbar\omega}\right),\nonumber\\
&\quad\approx \frac{s_0}{|E^{(s_0,j)}_\mathrm{3b}|/\hbar\omega}.
\end{align}
We therefore obtain the result for
\begin{align}\label{eq:c3deepbound}
C_3^{(s_0,j)}&\approx\frac{m|E_\mathrm{3b}^{(s_0,j)}|}{\hbar^2},\nonumber\\
&\approx\kappa_*^2e^{-2\pi j/s_0},\quad\quad\quad(E^{(s_0,j)}_\mathrm{3b}/\hbar\omega\to-\infty)
\end{align}
where we have used the result of Eq.~\eqref{eq:jnegative}.  The above result matches the operational definition $\left(\kappa_*\partial_{\kappa_*}E_\mathrm{3b}^{(s,j)}\right)|_a=-2\hbar^2C_3^{(s,j)}/m$ in the free-space limit \cite{PhysRevA.86.053633,PhysRevLett.106.153005}. 

\begin{figure*}[t!]
\centering
\includegraphics[width=14 cm]{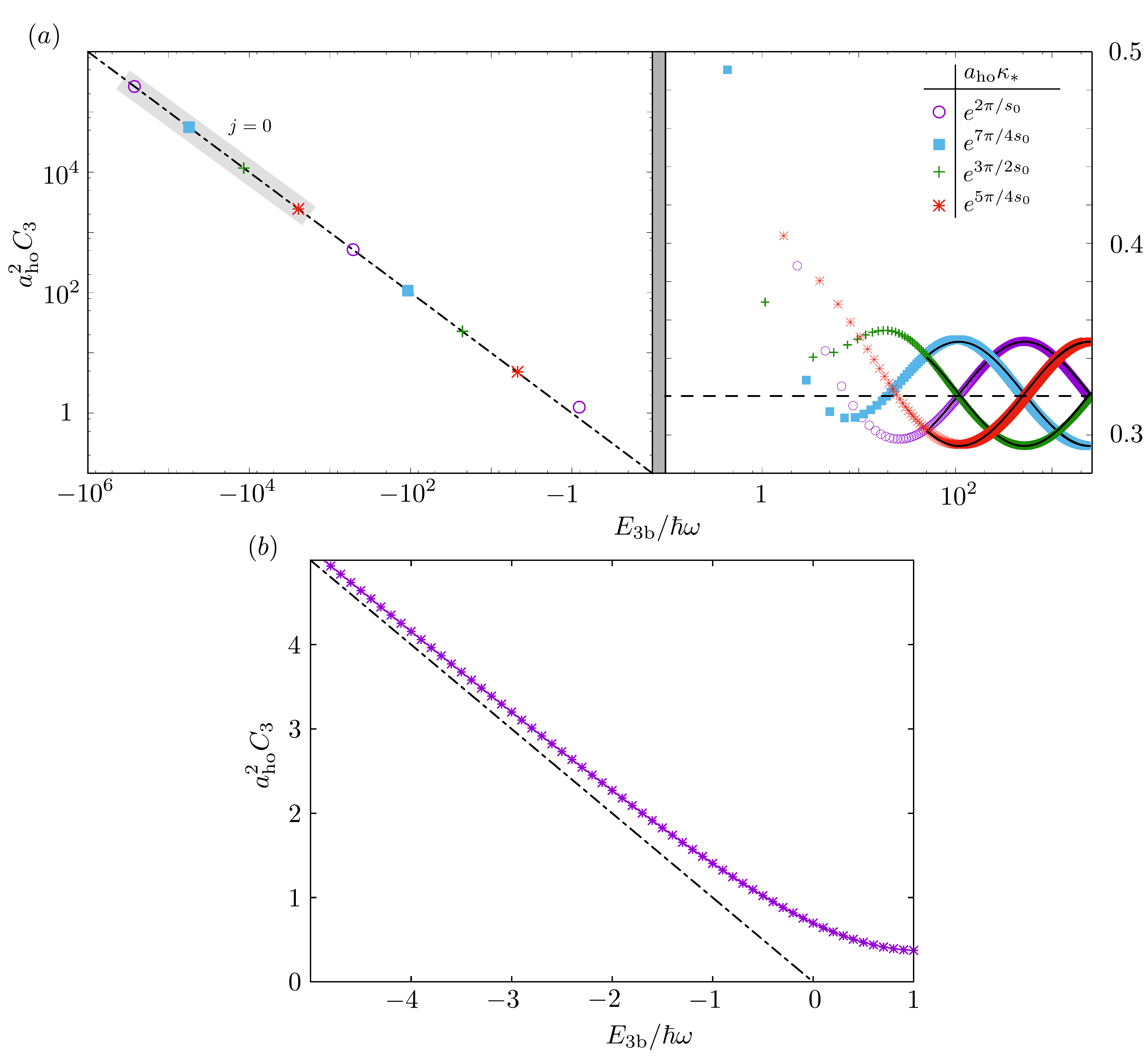}
\caption{Behavior of $a_\mathrm{ho}^2 C_3$ for both negative and positive eigenenergies in the Efimov channel.  (a) Results for four values of $a_\mathrm{ho}\kappa_*$ chosen evenly spaced over a single log-period as indicated in the key.  The $j=0$ indicates the ground-state trimer in the free-space limit in the sense of Eq.~\eqref{eq:jnegative} and are contained in the grey shaded region.  The black dashed-dotted line is the asymptotic result for $a_\mathrm{ho}^2C_3$ in the limit $E_\mathrm{3b}/\hbar\omega\to-\infty$ given by Eq.~\eqref{eq:c3deepbound}.  The solid black lines are the asymptotic results for $a_\mathrm{ho}^2C_3$ in the limit $E_\mathrm{3b}/\hbar\omega\to\infty$ given by Eq.~\eqref{eq:c3biglimit}, that oscillate around $a_\mathrm{ho}^2C_3\approx s_0/\pi$ indicated by the black dashed line.  (b) The transition of $a_\mathrm{ho}^2 C_3$ given by Eq.~\eqref{eq:c3trap} as a function of $E_\mathrm{3b}/\hbar\omega$ from positive to negative energies.  The black dashed-dotted line is the same as in (a).}\label{fig:c3}
\end{figure*}
\subsection{Results}

{\bf Case II  $(E_\mathrm{3b}/\hbar\omega\to+\infty)$:}  In this limit, the asymptotic expansion of the digamma function (Eq.~\eqref{eq:psilarge}) should be used carefully because we are taking the limit $|\arg z|\to \pi$.  Instead we use the reflection formula for the digamma function \cite{abramowitz1964handbook} to obtain
\begin{widetext}
\begin{equation}
\mathrm{Im}\ \psi\left(\frac{1-E_\mathrm{3b}^{(s_0,j)}/\hbar\omega+is_0}{2}\right)=-\pi\ \mathrm{Im}\ \cot\left(\frac{\pi}{2}(1+is_0-E_\mathrm{3b}^{(s_0,j)}/\hbar\omega)\right)+\mathrm{Im}\ \psi\left(\frac{1+E_\mathrm{3b}^{(s_0,j)}/\hbar\omega+is_0}{2}\right).
\end{equation} 
\end{widetext}
From the discussion of Case I, we know that the final term on the right in the above equation will vanish in the limit $E_\mathrm{3b}/\hbar\omega\to+\infty$.  Therefore, if the term involving the imaginary part of the cotangent is nonvanishing, it is the only term that needs to be considered.   Following this reasoning and using properties of cotangent for imaginary argument \cite{abramowitz1964handbook}, we find that
\begin{align}
-\pi\ \mathrm{Im}&\ \cot\left(\frac{\pi}{2}(1+is_0-E_\mathrm{3b}^{(s_0,j)}/\hbar\omega)\right)\nonumber\\
&=\frac{\pi\sinh(\pi s_0)}{\cosh(\pi s_0)+\cos(\pi E_\mathrm{3b}^{(s_0,j)}/\hbar\omega))},
\end{align}
which is nonvanishing.  We now take the limit $(E_\mathrm{3b}/\hbar\omega\to+\infty)$ and use Eq.~\ref{eq:jpos} to find
\begin{widetext}
\begin{align}
a_\mathrm{ho}^2 C_3^{(s_0,j)}&\approx\frac{s_0}{\pi}\frac{\cosh(\pi s_0)+\cos(s_0\ln(jR_t^2/a_\mathrm{ho}^2)-2\arg\Gamma(1+is_0))}{\sinh(\pi s_0)},\nonumber\\
&\approx\frac{s_0}{\pi}\left[1+0.085\cos(s_0\ln(jR_t^2/a_\mathrm{ho}^2)-2\arg\Gamma(1+is_0))\right].\quad\quad\quad(E_\mathrm{3b}/\hbar\omega\to\infty)\label{eq:c3biglimit}
\end{align}
\end{widetext}
The leading order result $a_\mathrm{ho}^2C_3^{(s_0,j)}\approx s_0/\pi$ was  obtained previously in Ref.~\cite{wernerthesis} for an analogous expression for the decay rate $\Gamma$ in this limit.  Having outlined the limiting behavior of $C_3^{(s_0,j)}$ in Sec.~\ref{sec:c3limiting}, we now discuss its behavior in general and over a range of $a_\mathrm{ho}\kappa_*$ as shown in Fig.~\ref{fig:c3}.  We find that $C_3^{(s_0,j)}$ matches the limiting forms (Eqs.~\eqref{eq:c3deepbound}, \eqref{eq:c3biglimit}) by $|E^{(s_0,j)}_\mathrm{3b}|/\hbar\omega\sim10^2$ as shown in Fig.~\ref{fig:c3}(a).  At intermediate energies, as $a_\mathrm{ho}$ decreases Efimov trimers whose free-space energies are comparable to $\hbar\omega$ are shifted above the zero-point energy of the trap as studied in Ref.~\cite{PhysRevLett.121.023401}.  This process repeats log-periodically as $a_\mathrm{ho}\kappa_*\to 0$.  A single log-period is shown in Fig.~\ref{fig:c3}(a), where the shift of the $j=2$ Efimov trimer is clearly visible.  For $|E_\mathrm{3b}|/\hbar\omega<1$, we find that $a_\mathrm{ho}^2C_3$ as a function of $E_\mathrm{3b}/\hbar\omega$ interpolates smoothly between the limiting behaviors as shown in Fig.~\ref{fig:c3}(b).   

In general we find that $C_3^{(s_0,j)}$ for a particular Efimov trimer is increased compared to the free space limit [see Eq.~\eqref{eq:c3deepbound}].  We echo the conclusion of Ref.~\cite{PhysRevA.97.033621} that the restriction of a trimer to a reduced region of space should correspond to an increase in the probability to find three bosons at short distances as encoded in $C_3^{(s_0,j)}$.   

Additionally, the phase of the log-periodic oscillation of $C_3^{(s_0,j)}$ for $E_\mathrm{3b}^{(j)}/\hbar\omega\gg1$ appears to adjust.  To see this behavior as a function of $a_\mathrm{ho}\kappa_*$, we rewrite Eq.~\ref{eq:c3biglimit} as
\begin{widetext}
\begin{equation}
a_\mathrm{ho}^2 C_3^{(s_0,j)}\approx\frac{s_0}{\pi}\left[1+0.085\cos(s_0\ln(j)-2s_0\ln(\kappa_*a_\mathrm{ho})+\Lambda)\right],\quad\quad\quad(E_\mathrm{3b}/\hbar\omega\to\infty)\label{eq:c3biglimit2}
\end{equation}
\end{widetext}
where the phase factor $\Lambda\approx0.697$.  The factor $2s_0\ln(\kappa_*a_\mathrm{ho})=2s_0\ln(\kappa_*a_\mathrm{ho}e^{\pi/s_0})$ $\mod 2\pi$, which gives the log-periodic behavior seen in Fig.~\ref{fig:c3}.   

\section{Two-body contact $C_2$}\label{sec:c2}
In this section, we derive the extensive two-body contacts $C_2$ for each trapped three-body eigenstate.  Analogous to the treatment in Sec.~\ref{sec:c3}, we begin from the definition of the non-normalized pair correlation function \cite{cohen2011advances}
\begin{equation}\label{eq:g2def}
G^{(2)}({\bf r}_1,{\bf r}_2)\equiv\langle \hat{\psi}^\dagger({\bf r}_1)  \hat{\psi}^\dagger({\bf r}_2) \hat{\psi}({\bf r}_2) \hat{\psi}({\bf r}_1)\rangle.
\end{equation}
For three bosons in vaccum, this is equivalent in first quantization to 
\begin{equation}
G^{(2)}({\bf r}_1,{\bf r}_2)=3!\int d^3 r_3 |\Psi({\bf r}_1,{\bf r}_2,{\bf r}_3)|^2. \label{eq:fewbodylink}
\end{equation}
We then integrate over the center of mass coordinate ${\bf c}\equiv({\bf r}_1+{\bf r}_2)/2$, and do a change of variables $\{{\bf c},{\bf r}_3\}\to\{{\bf C},{\bf r}_{3,12}\}$ where ${\bf r}_{3,12}\equiv{\bf r}_3-{\bf c}$.  The Jacobian for this transformation is unity, and we obtain
\begin{equation}
\int d^3c\ G^{(2)}({\bf r}_1,{\bf r}_2)=3!\int d^3 r_{3,12}|\psi({\bf r},{\bf r}_{3,12})|^2.
\end{equation}  
In the limit $|{\bf r}|\to0$, the left hand side of the above equation is related to $C_2$ \cite{PhysRevA.86.053633}, and therefore we find the following relation between the relative three-body wave function and the extensive two-body contact
\begin{equation}\label{eq:c2step}
3!\int d^3 r_{3,12} |\psi({\bf r},{\bf r}_{3,12})|^2\underset{{|\bf r|}\rightarrow 0}{=}\frac{C_2}{(4\pi)^2 r^2}.  
\end{equation}
In order to take the above limit, we rewrite the three-body relative wave function as
\begin{equation}
\Psi({\bf r},{\bf r}_{3,12})=\frac{\chi_0({\bf r},\boldsymbol{\rho})}{r\rho}+\frac{\chi_0({\bf r}^{(2)},\boldsymbol{\rho}^{(2)})}{r^{(2)}\rho^{(2)}}+\frac{\chi_0({\bf r}^{(3)},\boldsymbol{\rho}^{(3)})}{r^{(3)}\rho^{(3)}}
\end{equation}
where the functions $\chi_0$ have the important property of being finite for $|{\bf r}|\to 0$ and vanishing as $|\boldsymbol{\rho}|\to0$ \cite{0034-4885-80-5-056001}.  The vector superscripts indicate different Jacobi trees interrelated through the kinematic rotations
\begin{align}
{\bf r}^{(j)}&=-{\bf r}^{(i)}\cos\gamma^{(k)}+\boldsymbol{\rho}^{(i)}\sin\gamma^{(k)},\nonumber\\
\boldsymbol{\rho}^{(j)}&=-{\bf r}^{(i)}\sin\gamma^{(k)}-\boldsymbol{\rho}^{(i)}\cos\gamma^{(k)},\nonumber\\
\gamma^{(k)}&=\epsilon_{ijk}\pi/3,
\end{align}
where we have implicitly defined ${\bf r}\equiv{\bf r}^{(1)}$, $\boldsymbol{\rho}\equiv\boldsymbol{\rho}^{(1)}$.  With some rearrangement and the help of the kinematic rotations, we take the limit in Eq.~\eqref{eq:c2step} to obtain 
\begin{widetext}
\begin{align}
C_2&=3! (4\pi)^2\lim_{|r|\to0} \int d^3 r_{3,12}\ r^2\left|\frac{\chi_0({\bf r},\boldsymbol{\rho})}{r\rho}+\frac{\chi_0({\bf r}^{(2)},\boldsymbol{\rho}^{(2)})}{r^{(2)}\rho^{(2)}}+\frac{\chi_0({\bf r}^{(3)},\boldsymbol{\rho}^{(3)})}{r^{(3)}\rho^{(3)}}\right|^2,\nonumber\\
&=3!(4\pi)^2\int d^3r_{3,12}\left|\frac{\chi_0(0,\boldsymbol{\rho})}{\rho}\right|^2.
\end{align}
\end{widetext}
Written in terms of the trapped eigenstates outlined in Sec.~\ref{sec:background}, this becomes
\begin{equation}\label{eq:c2expression}
C_2^{(s,j)}=72\pi^2|\mathcal{N}_{s,j}|^2|\sin(s\pi/2)|^2A_{j,j}^{s,s},
\end{equation}
where we have indexed the extensive two-body contact specific to the eigenstate $(s,j)$ as $C_2^{(s,j)}$.  The notation $A_{j,j}^{s,s}$ is shorthand for diagonal components of the array
\begin{equation}\label{eq:atensor}
A^{s,s'}_{j,j'}\equiv\sqrt{\frac{3}{2}}\int dz F_j^{s}(z)[F^{(s')}_{j'}(z)]^*.
\end{equation}

It is possible to obtain analytic expressions for the array $A_{j,j'}^{s,s'}$ using the trapped hyperradial eigenstates outlined in Sec.~\ref{sec:background}.  We are concerned however only with the components with $j=j'$ and $s=s'$ required to evaluate $C_2^{(s,j)}$ via Eq.~\eqref{eq:c2expression}.  For $s=is_0$, Eq.~\ref{eq:atensor} is an integral over Whittaker functions that can be found in a table \cite{gradshteyn2007table} with result
\begin{widetext}
\begin{align}
A^{s_0,s_0}_{j,j}&=\sqrt{\frac{3}{2}}a_\mathrm{ho}\mathrm{Re}\left[\frac{\Gamma[1/2]\Gamma[1/2+is_0]\Gamma[-is_0]}{\Gamma\left[\frac{1-is_0-E^{(s_0,j)}_\mathrm{3b}/\hbar\omega}{2}\right]\Gamma\left[\frac{2+is_0-E^{(s_0,j)}_\mathrm{3b}/\hbar\omega}{2}\right]}\right. \nonumber\\
&\left. \times\text{}_3F_2\left(\frac{1}{2},\frac{1}{2}+is_0,\frac{1-E^{(s_0,j)}_\mathrm{3b}/\hbar\omega+is_0}{2};1+is_0,\frac{2-E^{(s_0,j)}_\mathrm{3b}/\hbar\omega+is_0}{2};1\right)          \right],\label{eq:aefimov}
\end{align}
\end{widetext}
where $\text{}_3F_2$ is a generalized hypergeometric function.  We note that an equivalent expression to Eq.~\ref{eq:aefimov} was first obtain in Ref.~\cite{wernerthesis} for the loss-rate $\Gamma$.  For channels with $s^2>0$, Eq.~\ref{eq:atensor} is an integral over generalized Laguerre polynomials that can be recast using the recurrence relation \cite{abramowitz1964handbook}
\begin{equation}\label{eq:laguerrerelation}
L_{j}^{(s)}(z)=\sum_{l=0}^{j'}\begin{pmatrix}
(2s-1)/2+j-l\\
j-l\\
\end{pmatrix} L_l^{\left(-1/2\right)}(z),
\end{equation}
where the $\left(\cdot\right)$ is the generalized binomial coefficient.  Using this relation, the array $A^{s,s}_{j,j}$ can be found tabulated in Ref.~\cite{gradshteyn2007table}, and we find that
\begin{widetext}
\begin{align}
A^{s,s}_{j,j}=\frac{1}{2}\sqrt{\frac{3}{2}}a_\mathrm{ho}\left(\frac{2s-1}{2}\right)!
\begin{pmatrix}
-1/2\\
j\\
\end{pmatrix}
\begin{pmatrix}
s+j-1/2\\
j\\
\end{pmatrix}
\text{}_3F_2\left(-j,-j-s,1/2;-j-s+1/2,-j+1/2;1\right)
.\label{eq:Aunivs2}
\end{align}
\end{widetext}
In the following subsection, we derive some limiting cases of $A^{s,s}_{j,j}$, and then discuss results for $\mathcal{C}_2^{(s,j)}$ in general.


\subsection{Limiting cases}\label{sec:limitc2}

Due to the $\text{}_3F_2$ function in Eqs.~\eqref{eq:aefimov}, \eqref{eq:Aunivs2} it is unclear how to derive limiting expressions for $C_2^{(s,j)}$ in the limit $j\to+\infty$ \cite{olver2010nist}.  It is however possible to derive expressions in the opposite limits, which includes the $j\to-\infty$ limit in the Efimov channel and the case $j=0$ for the universal channels.  

For $s=is_0$ in the $j\to-\infty$ limit, the $\text{}_3F_2$ function can be expanded up to order $(E_\mathrm{3b}^{(s_0,j)})^{-m}$ where $m$ is any nonnegative integer \cite{olver2010nist}.  Alternatively, using the asymptotic form of the hyperradial wave function in this limit (Eq.~\ref{eq:psiefimov}), the relevant integral for $A_{j,j}^{s_0,s_0}$ can be found tabulated in Ref.~\cite{gradshteyn2007table}, and we obtain 
\begin{equation}\label{eq:psiefimov2}
A_{j,j}^{s_0,s_0}\underset{{j}\rightarrow-\infty}{=}a_\mathrm{ho}\frac{\sqrt{3}\pi^2}{8\sqrt{|E_\mathrm{3b}^{(s_0,j)}|/\hbar\omega}\cosh\pi s_0},
\end{equation}
which gives the following expression for $C_2^{(s,j)}$
\begin{widetext}
\begin{align}
a_\mathrm{ho}C_2^{(s_0,j)}\underset{{j}\rightarrow-\infty}{=}&144\pi^2\sqrt{|E_\mathrm{3b}^{(s_0,j)}|/\hbar\omega}\frac{\sinh(\pi s_0/2)\tanh(\pi s_0)}{9\pi s_0\sinh(\pi s_0/2)-8\sqrt{3}\pi\cosh(\pi s_0/6)+18\cosh(\pi s_0/2)}, \nonumber\\
C_2^{(s_0,j)}\approx&53.097\kappa_* e^{-\pi j/s_0}.\quad\quad\quad(E_\mathrm{3b}/\hbar\omega\to-\infty)\label{eq:c2efimov}
\end{align}
\end{widetext}
This matches the result in Ref.~\cite{PhysRevA.83.063614} for $C_2$ for an Efimov trimer obtained via an alternative approach by considering the high-momentum tail of the single-particle momentum distribution.  

For $s^2>0$ with $j=0$, the array $A^{s,s}_{0,0}$ has the form
\begin{equation}  
A_{0,0}^{s,s}=a_\mathrm{ho}\frac{1}{2}\sqrt{\frac{3}{2}}(s-1/2)!,
\end{equation}
and we obtain the following expression
\begin{widetext}
\begin{align}
a_\mathrm{ho}C_2^{(s,0)}&=144\sqrt{2}\pi\frac{\Gamma(s+1/2)}{\Gamma(s)}\frac{\sin(\pi s/2)}{8\sqrt{3}\pi\cos(\pi s/6)-18\cos(\pi s/2)+9\pi s\sin(\pi s/2)},\\
&\underset{{s}\rightarrow{+\infty}}{=}16\sqrt{2}s^{-1/2},\label{eq:c2jzero}
\end{align}
\end{widetext}
which vanishes in the limit $s\to+\infty$.

\begin{figure*}[t!]
\centering
\includegraphics[width=14 cm]{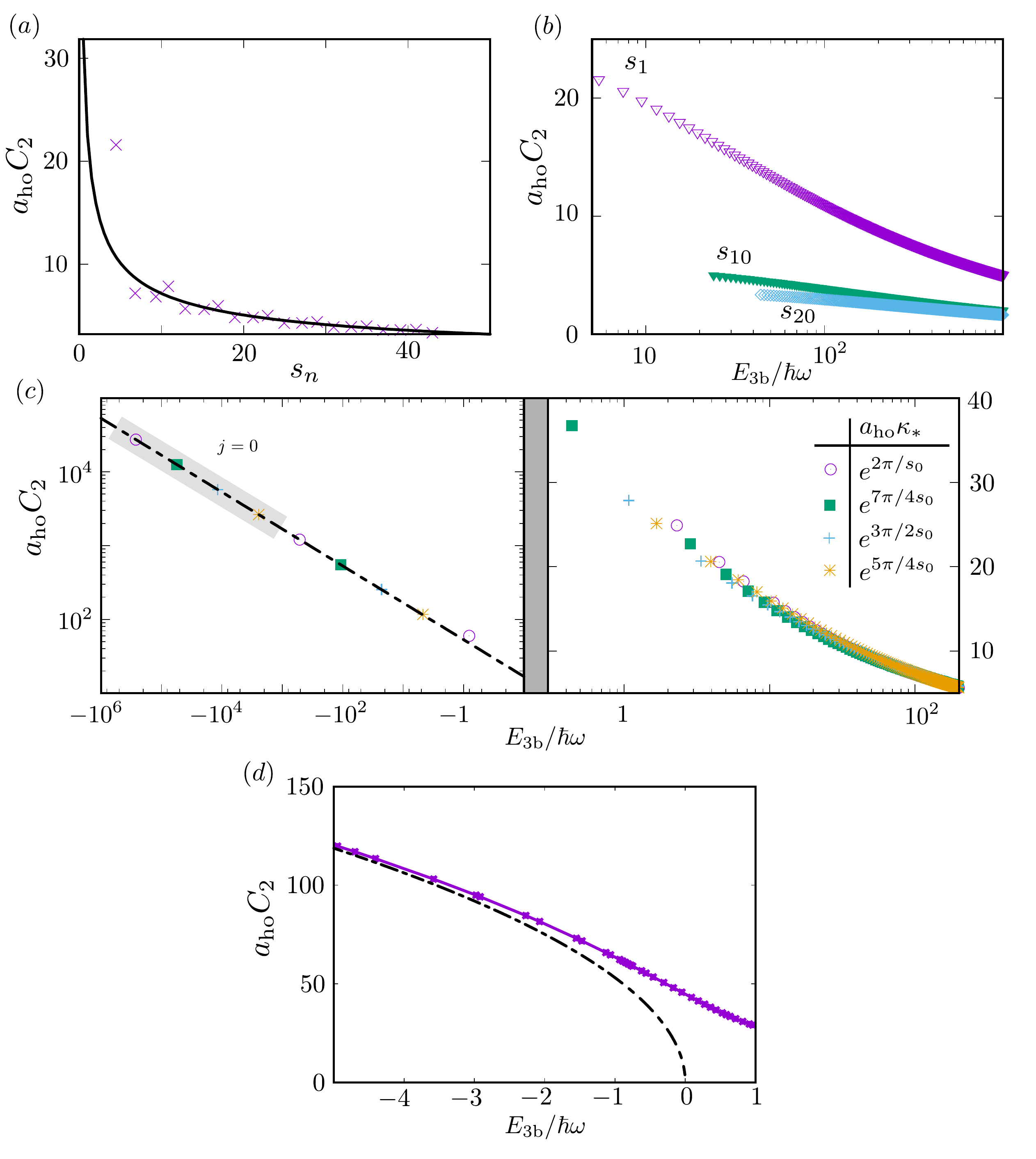}
\caption{Behavior of $a_\mathrm{ho} C_2$ for both negative and positive eigenenergies in the universal and Efimov channels. (a) The behavior of $a_\mathrm{ho}C_2^{(s,0)}$ for $s^2>0$ compared against the $s\to+\infty$ asymptotic result (black solid line) Eq.~\eqref{eq:c2jzero}.  (b)  Results for $a_\mathrm{ho}C_2^{(s,j)}$ for $s^2>0$  over a range of energies in the three universal channels $s_1\approx4.465$, $s_{10}\approx22.932$, and $s_{20}\approx42.967$. (c) Results for $a_\mathrm{ho}C_2^{(s_0,j)}$ for four values of $a_\mathrm{ho}\kappa_*$ chosen evenly spaced over a single log-period (same as in Fig.~\ref{fig:c3}) as indicated in the key.  The black dashed-dotted line is the asymptotic result for $a_\mathrm{ho}C_2$ in the limit $E_\mathrm{3b}/\hbar\omega\to-\infty$ given by Eq.~\eqref{eq:c2efimov}.  The grey shaded region indicates the $j=0$ eigenstates. (d) The transition of $a_\mathrm{ho} C_2^{s_0,j}$ as a function of $E_\mathrm{3b}/\hbar\omega$ from positive to negative energies.  The black dashed-dotted line is the same as in (c).}\label{fig:c2}
\end{figure*}
\subsection{Results}

Having outlined some of the limiting behaviors of $C_2^{(s,j)}$ in Sec.~\ref{sec:limitc2}, we now analyze its behavior in general over a range of universal channels and for various values of $a_\mathrm{ho}\kappa_*$ in the Efimov channel as shown in Fig.~\ref{fig:c2}.  First we discuss results for the universal channels shown in Fig.~\ref{fig:c2}(a-b).  In Fig.~\ref{fig:c2}(a), we see that our asymptotic prediction for $s\to+\infty$ with $j=0$ given by Eq.~\eqref{eq:c2jzero} matches our calculation of $a_\mathrm{ho}C_2$ as $s_n\gg1$ (solid black line).  At larger $j$, the behavior of $a_\mathrm{ho}C_2$ is shown in Fig.~\ref{fig:c2}(b), where we see a gradual decay.  The $j\to+\infty$ asymptotic form of $a_\mathrm{ho}C_2$ was not derived in Sec.~\ref{sec:limitc2}, although we estimate its form by fitting $a_\mathrm{ho}C_2$ to a power law $j^{-\alpha}$.  We find that for extremely large energies $E_\mathrm{3b}/\hbar\omega>10^6$ this exponent approaches $\alpha\approx 0.5$ in both universal and Efimov channels.  This power law can be motivated by approximating $1/2-j-s\approx -j-s$ in the arguement of the $\text{}_3F_2$ function in Eq.~\ref{eq:Aunivs2}.  The  $\text{}_3F_2$ function then reduces to $\text{}_2F_1$ \cite{olver2010nist} and the $j\to+\infty$ limit can be taken to obtain a $j^{-1/2}$ power law scaling for $a_\mathrm{ho}C_2$.

Our results for $a_\mathrm{ho}C_2$ in the Efimov channel are shown in Fig.~\ref{fig:c2}(c-d) for both positive and negative energies starting at $j=0$.  We see that the $E_\mathrm{3b}/\hbar\omega\to-\infty$ asymptotic form given in Eq.~\eqref{eq:c2efimov} (dot-dashed black line) matches our calculation of $a_\mathrm{ho}C_2$ as shown in Fig.~\ref{fig:c2}(c).  At positive energies, a gradual decay of $a_\mathrm{ho}C_2$ is evident as in the universal case [Fig.~\ref{fig:c2}(c)], and the log-periodic oscillations that were prominent in $C_3$ in Fig.~\ref{fig:c3} are absent.  For varying $a_\mathrm{ho}\kappa_*$, our results follow the same trends.  As $j\to+\infty$ we observe numerically the same $j^{-1/2}$ power discussed for the universal channels.  Additionally, in \ref{fig:c2}(d) we show the behavior of $a_\mathrm{ho}C_2^{s_0,j}$ as a function of $E_\mathrm{3b}/\hbar\omega$ in the region $|E_\mathrm{3b}|/\hbar\omega<1$.  Here we find, as in Fig.~\ref{fig:c3}(b), a smooth interpolation between limiting behaviors.

In general, we find that over the same energy ranges $a_\mathrm{ho}C_2$ is larger in the Efimov channel than in the universal channels.  Physically, we understand this as a larger probability for finding correlated pairs of atoms at short-distances, which is measured by $C_2$ \cite{PhysRevA.78.053606}.  This should be larger in the absence of a repulsive hyperradial barrier for $s^2>0$ (see Eq.~\eqref{eq:hyper}.)  As $s$ increases, this barrier becomes more repulsive [see Eq.~\eqref{eq:hyper}], and it is therefore less likely to find correlated pairs at short-distances.  This is reflected in Fig.~\ref{fig:c2}(a-b).  When comparing to the free-space result in Eq.~\eqref{eq:c2efimov} for an Efimov trimer, we find that the trap leads to a relative increase of $C_2$.  Tighter traps increasingly confine the atoms, and therefore it makes sense that the probability to find correlated pairs should increase accordingly.

\section{Discussion}\label{sec:discussion}
\begin{table*}[ht!]
\caption{Comparison of the results for $E_\mathrm{3b}/\hbar\omega$, $a_\mathrm{ho}C_2$, and $a_\mathrm{ho}^2C_3$ in this work with the results of Ref.~\cite{PhysRevA.97.033621} for the energetically lowest eight eigenstates.  To obtain these results, we have chosen an oscillator length $a_\mathrm{ho}\kappa_*\approx 0.43$.}
\label{table:C2}
\centering
\begin{tabular}{l l c c c c c c}
\hline
\hline
\noalign{\smallskip}
$s$ & $j$ & $E_\mathrm{3b}/E_\mathrm{ho}$ \cite{PhysRevA.97.033621}& $E_\mathrm{3b}/E_\mathrm{ho}$ & $a_\mathrm{ho}C_2$ \cite{PhysRevA.97.033621} & $a_\mathrm{ho}C_2$ & $a_\mathrm{ho}^2C_3$ \cite{PhysRevA.97.033621} & $a_\mathrm{ho}^2C_3$ \\
\noalign{\smallskip}
\hline\hline
\noalign{\smallskip}
$is_0$ & $0$ &  0.5000 & 0.5000 & 35.6 & 35.7 & 0.478 & 0.469\\
\noalign{\smallskip}
$is_0$ & $1$ &  2.9217 & 2.9019 & 22.4 & 22.5 & 0.334 & 0.327\\
\noalign{\smallskip}
$is_0$ & $2$ &  5.0895 & 5.0587 & 18.8 & 19.0 & 0.323 & 0.313\\
\noalign{\smallskip}
$s_1$ & $0$ &  5.5187 & 5.4653 & 21.6 & 21.6 & $9\times10^{-6}$ & 0\\
\noalign{\smallskip}
$is_0$ & $3$ &  7.2005 & 7.1597 & 17.3 & 17.0 & 0.326 & 0.310\\
\noalign{\smallskip}
$s_1$ & $1$  & 7.5258 & 7.4653 & 20.5 & 20.6 & $3\times10^{-5}$ & 0\\
\noalign{\smallskip}
$s_2$ & $0$ & 7.8452 & 7.8184 & 7.15 & 7.17&$2\times10^{-7}$ & 0\\
\noalign{\smallskip}
$is_0$ & $4$ &  9.2874 & 9.2351 & 15.4 & 15.7 & 0.336 & 0.311\\
\hline
\end{tabular}
\end{table*}
In this work, analytic solutions of the trapped unitary three-boson problem from Ref.~\cite{PhysRevLett.97.150401} are utilized to derive analytic expressions for the two and three-body contacts for each trapped three-body eigenstate.  These results can be used along with the set of universal relations associated with the contacts~\cite{PhysRevLett.106.153005,PhysRevA.86.053633} to predict, for instance, the high-frequency tail of the rf transition rate, the virial theorem, loss-rates, and total energy.  We derive the large energy asymptotics of the contacts, and compare with existing results in the literature when possible finding good agreement.  At intermediate energies, the only results in the literature that we are aware of are from Ref.~\cite{PhysRevA.97.033621}, which were obtained using more realistic Gaussian potentials with a finite-range.  We take a wave function based approach to obtain the contacts from the short-distance behavior of the pair and triplet correlation functions.  In Ref.~\cite{PhysRevA.97.033621}, however, the contacts were obtained from their operational definitions~\cite{PhysRevLett.106.153005,PhysRevA.86.053633}
\begin{align}
 \left(a\partial_{a}E_\mathrm{3b}^{(s,j)}\right)|_{\kappa_*}&=\frac{\hbar^2}{8\pi m a}C_2^{(s,j)}, \\
  \left(\kappa_*\partial_{\kappa_*}E_\mathrm{3b}^{(s,j)}\right)|_a&=-\frac{2\hbar^2}{m}C_3^{(s,j)}, 
  \end{align}
by taking the derivatives of three-body eigenenergies with variations of $a$ and $\kappa_*$.  This amounted to changing the underlying parameters of the finite-range Gaussian potentials and approximating the derivatives via finite differencing.  In Table~\ref{table:C2}, we compare against results in that work, finding good agreement for the eigenenergies and contacts to within a few percent or less.  In Ref.~\cite{PhysRevA.97.033621}, the convergence of eigenenergies was estimated as $10^{-3}E_\mathrm{ho}$ or better.  The eigenenergies calculated in the present work can be calculated from Eq.~\eqref{eq:spectrum} to arbitrary precision.  We attribute the disagreement therefore to the absence of finite-range effects in the zero-range model in the present work.       

Finally, we note that the method outlined in these notes to obtain $C_2$ and $C_3$ through the limiting behavior of the correlation functions applies also away from unitarity at finite scattering lengths provided the zero-range approximation can be made \cite{Portegies2011}.  Additionally, the derivation in of $C_2$ from a three-body model presented in Sec.~\ref{sec:c2} may be generalized in the spirit of Ref.~\cite{PhysRevLett.120.100401} to study the dynamics of the two-body contact following an interaction quench.  We address this latter problem elsewhere \cite{colussic3therm} and leave the former as the subject of future study.

{\it Acknowledgements.}  We acknowledge discussions with John Corson, Jos\'e D'Incao, and Servaas Kokkelmans.  This research was funded by the Nederlandse Organisatie voor Wetenschappelijk Onderzoek (NWO) grant number 680-47-623. 

\bibliographystyle{apsrev4-1}
\bibliography{references.bib}

\end{document}